\documentclass[conference]{IEEEtran}

\usepackage{cite}
\usepackage{amsmath,amssymb,amsfonts}
\usepackage{algorithmic}
\usepackage{graphicx}
\usepackage{textcomp}
\usepackage{xcolor}
\usepackage{threeparttable}
\usepackage{multirow}
\usepackage[linesnumbered,lined,boxed,vlined]{algorithm2e}
\usepackage{bm}
\usepackage{booktabs}
\usepackage{subfig}
\usepackage{array}
\usepackage{enumerate}
\usepackage{fancyhdr}
\def\BibTeX{{\rm B\kern-.05em{\sc i\kern-.025em b}\kern-.08em
    T\kern-.1667em\lower.7ex\hbox{E}\kern-.125emX}}

\fancypagestyle{firstpage}{

  \fancyhead[C]{This paper is published in 2024 Design, Automation and Test in Europe Conference (DATE'24)} 
  \fancyfoot[C]{}
}

\begin{document}

\title{PIMSYN: Synthesizing Processing-in-memory \\ CNN Accelerators
}

\author{
    \IEEEauthorblockN{Wanqian Li, Xiaotian Sun, Xinyu Wang, Lei Wang, Yinhe Han, Xiaoming Chen\IEEEauthorrefmark{1}}
    \IEEEauthorblockA{Institute of Computing Technology, Chinese Academy of Sciences
    \\University of Chinese Academy of Sciences
    \\\IEEEauthorrefmark{1}Corresponding author: chenxiaoming@ict.ac.cn}
}

\maketitle

\thispagestyle{firstpage}
\begin{abstract}
    Processing-in-memory architectures have been regarded as a promising solution for CNN acceleration. Existing PIM accelerator designs rely heavily on the experience of experts and require significant manual design overhead. Manual design cannot effectively optimize and explore architecture implementations. In this work, we develop an automatic framework PIMSYN for synthesizing PIM-based CNN accelerators, which greatly facilitates architecture design and helps generate energy-efficient accelerators. PIMSYN can automatically transform CNN applications into execution workflows and hardware construction of PIM accelerators.   To systematically optimize the architecture, we embed an architectural exploration flow into the synthesis framework, providing a more comprehensive design space. Experiments demonstrate that PIMSYN improves the power efficiency by several times compared with existing works. PIMSYN can be obtained from https://github.com/lixixi-jook/PIMSYN-NN.
\end{abstract}


\section{Introduction}

    In recent years, processing-in-memory (PIM) has widely been studied for convolution neural network (CNN) inference for performance and power efficiency improvements, which remarkably reduces data accesses between arithmetic and storage components (e.g., \cite{atomlayer, ISAAC, pipelayer, prime,puma}). Compared with traditional CNN accelerators, PIM-based accelerators improve the power efficiency by 2-3 orders of magnitude \cite{prime}. Plenty of works implement PIM accelerators based on crossbars composed of ReRAM devices. The crossbar structure can perform an in-situ analog matrix-vector multiplication (MVM) within $\mathcal{O}(1)$ time, which not only eliminates the data accesses for CNN weights but also exploits high parallelism of CNN computation.

    PIM-based CNN accelerator design has the following characteristics and challenges which are distinct from conventional CMOS-based CNN accelerators.
         1) \textbf{Inter-layer pipelining.} Instead of layer-by-layer execution and transferring both weights and activations for every layer in CMOS-based accelerators, a PIM accelerator stores \textcolor{black}{the entire CNN's} weights within ReRAM cells and fulfills inter-layer pipelining. PIM accelerators involve a concept of \textit{weight duplication}, which means storing a layer's weights for multiple copies. 
         Duplicating weights enhances computation throughput but greatly expands the design space of both hardware architecture and dataflow scheduling.       
         2) \textbf{Communication bottleneck.} PIM accelerators are typically organized by  macros (a macro is typically composed of crossbars and necessary peripheral components), interconnected via a network-on-chip (NoC) or bus. Although PIM alleviates weight access, the demand for communicating activations and intermediate results persists within and between macros. Particularly when weights are duplicated, as computation parallelism increasing, communication emerges as the bottleneck of the accelerator performance.
         3) \textbf{Energy-intensive peripheral components.} In PIM accelerators, peripheral components, like ADCs and DACs, consume over 60\% of the total power \cite{ISAAC}, playing a critical role in optimizing the power efficiency of PIM architectures.

    To develop power-efficient PIM-based CNN accelerators, these factors must be explored to comprehensively optimize power, performance and area (PPA), which greatly amplifies the difficulty of PIM architecture design. Existing PIM-based CNN accelerators are mostly manually designed,  which requires both experienced developers and substantial human efforts. 
    It is almost impossible to create PPA-optimized CNN accelerators depending solely on expert experience. Especially when the scale of the design space is extensive, existing works miss a thorough analysis about the mentioned characteristics. Therefore, developing EDA tools which can automatically generate PPA-optimized PIM accelerators is extremely useful.

    \textcolor{black}{
    There are a few works involving the generation and exploration of PIM-based CNN accelerators. Works~\cite{gibbon2,nacim} simply enumerate the architecture parameters (e.g., crossbar size, ReRAM resolution, etc.) assuming a fixed architecture organization and a determined dataflow. AutoDCIM~\cite{autodcim} optimizes solely at circuit-level and layout-level for digital PIM architectures, using a template-based generation approach. PIM-HLS~\cite{pim-hls} places more emphasis on solving memory distribution (SRAM and ReRAM) problems in heterogeneous architectures. But it lacks comprehensive optimization and exploration for homogeneous architectures. The limitation indicates that PIM-HLS cannot be directly applied to the more common practice based on the same device. None of the above succeeds in fulfilling a full-stack synthesis flow to automatically convert CNN tasks into homogeneous PIM architectures, as well as comprehensively considering accelerators' design space. 
    }

   \textcolor{black}{To address the above challenges, this work makes the following contributions.}
   \begin{itemize}
    \item We propose PIMSYN, a full-stack automatic synthesis framework for PIM-based CNN accelerators. PIMSYN realizes a one-click transformation from CNN applications to PIM architectures.
    \item Given the CNN tasks and user-defined power constraint, PIMSYN performs a series of synthesis steps in which design space exploration (DSE) is performed to generate power-efficient PIM accelerators. The design space covers exploration on both dataflows and architectures.
    \item By defining a set of PIM-friendly intermediate representations (IRs) to express various design choices, PIMSYN greatly expands the accelerator design space and increases the architectural optimization opportunities.
    \item Experiments demonstrate the superiority of PIMSYN.
   \end{itemize}

    \section{Preliminary}\label{sec2}

    \subsection{PIM-based CNN Acceleration}\label{sec:base}

\begin{figure}[t]
            \centering
            \includegraphics[width=.7\columnwidth]{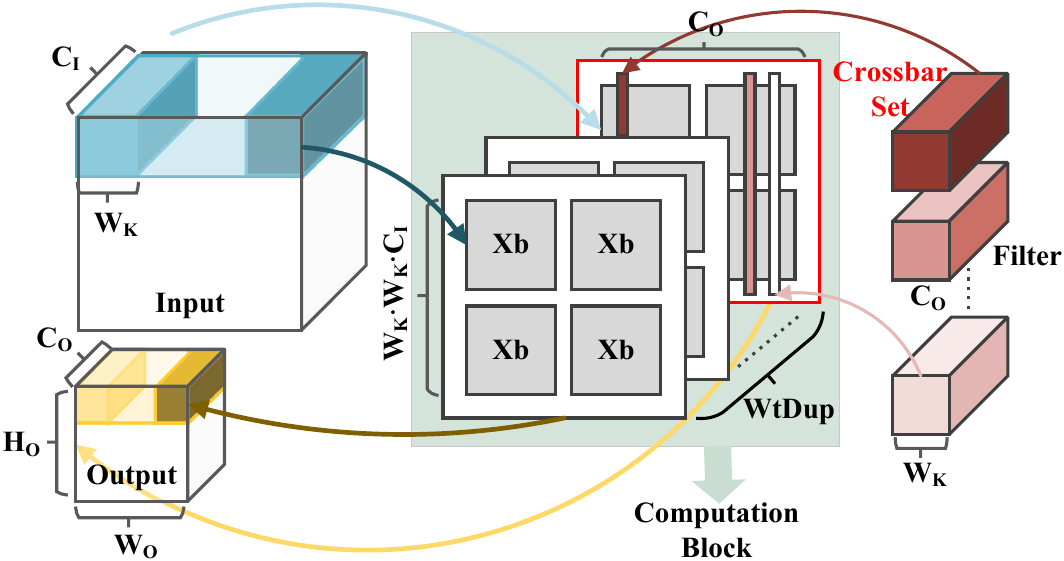}\\
            \caption{{}{Crossbar-accelerated convolution computation and weight duplication.}}\label{fig:cnn}
        \end{figure}

        Fig.~\ref{fig:cnn} illustrates the scheme of crossbar-based hardware accelerating convolutions. Weights from one kernel filter are programmed into the same column of one or more crossbars. A convolution layer need $C_O$  columns and ${W}_{K}\times {W}_{K}\times {C}_{I}$ rows in total. Due to the limited size of crossbars and resolution of devices, mapping a layer’s weights often needs multiple crossbars, denoted as a \textit{crossbar set}. The number of crossbars in one crossbar set can be calculated as
        \begin{equation}\label{eq:set}\footnotesize
          set=\left\lceil \frac{W_{K} W_{K}  C_{I}}{XbSize} \right\rceil \times \left\lceil \frac{C_{O}}{XbSize} \right\rceil \times \left\lceil \frac{PrecWt}{ResRram} \right\rceil,
        \end{equation}
        where ${XbSize}$, ${PrecWt}$ and ${ResRram}$ are the crossbar size, weight precision and ReRAM cell resolution, respectively. After loading ${W}_{K}\times {W}_{K}\times {C}_{I}$ inputs to crossbars' word lines, a crossbar set can calculate $C_O$ outputs in one step. 
         By duplicating the weights of a layer by $WtDup$ times, $WtDup\times C_O$ outputs can be computed in parallel. This computation process with weights duplicated by $WtDup$ times  is  denoted as a \textit{computation block}. 
        The inference performance is closely related to the weight duplication factors of  layers, namely, $\lceil W_{O}\times H_{O}/WtDup\rceil$ steps for completing a layer. 
        \textcolor{black}{If the input activation precision exceeds the DAC resolution, bit-level iterations will be introduced. Each iteration processes input bits equivalent to the DAC precision.}

    \subsection{Architecture Abstraction of PIM CNN Accelerators}
   
        \begin{figure}[t]
            \centering
            \includegraphics[width=.75\columnwidth]{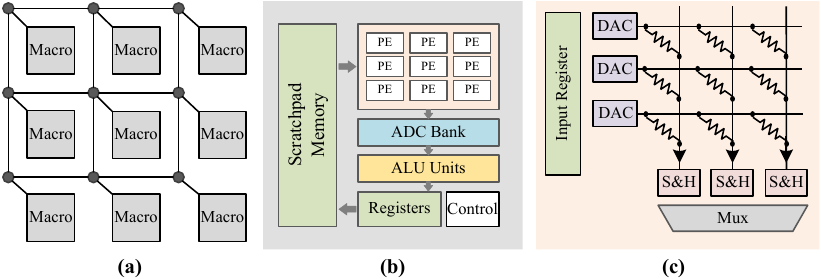}\\
            \caption{Architecture abstraction of PIM-based CNN accelerators. (a) Overall architecture. (b) Macro. (c) PE.}\label{fig:arch}
        \end{figure}

        \begin{figure*}[t]
            \centering
            \includegraphics[width=.7\textwidth]{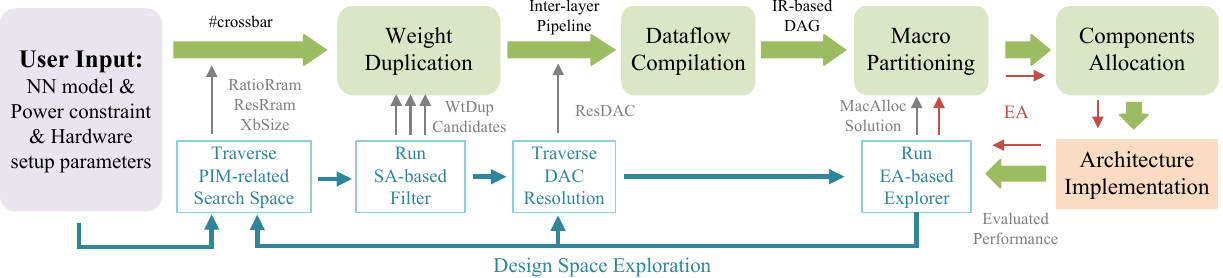}\\
            \caption{Overview of PIMSYN framework.}\label{fig:overview}
        \end{figure*}
    
        PIMSYN generates PIM-based CNN accelerators based on a PIM-oriented architecture abstraction (i.e., template), as shown in Fig.~\ref{fig:arch}. It is composed of a three-level macro-PE-crossbar hierarchy, where macros are interconnected through a NoC. Different layers are simultaneously executed in macros through pipelining. Communication between layers brings inter-macro synchronization. Each macro consists of a scratchpad memory, a PE array, an ADC bank, ALU components, register files and a controller. In PIMSYN, macros can be configured either identical or specialized for different layers. During CNN inference, PE  performs MVM operations and produces analog outputs. A PE includes input registers, DACs, a crossbar, sample and hold (S\&H) units and output multiplexers. Results from PE arrays are further converted by the ADC bank and calculated by ALUs which support vector operations (e.g.,  shift-and-add, pooling, ReLU, etc.).  The architecture abstraction is compatible with most previous works, and provides enough flexibility for various implementations. In PIMSYN, all the above components are configurable and parameterizable to enable a comprehensive DSE.

    \section{PIMSYN Framework Overview}

        Fig.~\ref{fig:overview} describes the overview of PIMSYN. The input of PIMSYN includes  a CNN model structure described in the ONNX format~\cite{onnx}, a total power constraint, and hardware setup parameters (e.g., ReRAM's, ADC's and DAC's latency and power). Through a series of synthesis stages together with the integrated DSE flow, PIMSYN  automatically generates the architecture of a PIM-based accelerator with maximized power efficiency. The generated solution also specifies the dataflow scheduling, i.e., when and where each computation task is performed. Since power constraint is an input, maximizing power efficiency is equivalent to maximizing performance. 
        \textcolor{black}{To be mentioned, the CNN model has well been designed, trained, and  quantified, and is an input of PIMSYN.}
         Hardware synthesis will not cause any accuracy loss for given CNN algorithms. To ensure that, in PIMSYN, we set the resolution of ADCs to satisfy the minimum resolution requirement according to \cite{ISAAC}.

       The synthesis process of PIMSYN is abstracted into four stages. 
     1)   \textbf{Weight duplication.} It decides the weight duplication factor for each layer. It determines layers' parallelism and initializes the inter-layer pipeline. 
      2)   \textbf{Dataflow compilation.} It transforms the CNN structural description into the proposed  IRs. Dataflow is described by an IR-based directed acyclic graph (DAG), depicting the execution flow and relations between operations. \textcolor{black}{Weight duplication can greatly influence the produced dataflow.} 
       3)   \textbf{Macro partitioning.} It distributes computation tasks of each layer to respective macros. It also determines data access patterns both within and between macros.
        4)  \textbf{Components allocation.} It assigns functional components inside each macro, such as ADCs, DACs and ALUs. After completing this stage, the accelerator's implementation details are finalized.
        Among \textcolor{black}{the four stages}, the first two stages optimize the dataflow, and the last two optimize the hardware implementation based on the dataflow. 

        \begin{table}[t]
          \caption{Design space of PIM-based CNN accelerators.}\label{tab:dse}\scriptsize
          \centering
          \setlength{\tabcolsep}{1pt}
          \begin{tabular}{|c|p{4.6cm}|c|c|}
          \hline
          Design variable       & \centering Definition  & Gibbon \cite{gibbon2}  & NACIM  \cite{nacim}  \\ 
          \hline
          $RatioRram$          & Ratio of all ReRAMs' power to total power, ranging from 0.1 to 0.4& No     & No   \\ 
          \hline
          $\bm{WtDup}$     & ${WtDup}^{i}$, layer $i$'s weight duplication factor & No     & No  \\ 
          \hline
          $XbSize$             & Size of crossbar, like 128, 256, 512  & Yes    & Yes     \\ 
          \hline
          $ResRram$            & ReRAM resolution, like 1, 2, 4& Yes    & Yes   \\ 
          \hline
          $ResDAC$             & DAC resolution, like 1, 2, 4 & Yes    & Yes   \\ 
          \hline
          $\bm{MacAlloc}$  & ${MacAlloc}^{i}$, \textcolor{black}{\# of macros assigned to layer $i$}  & No     & Yes   \\ 
          \hline
          $\bm{CompAlloc}$ & ${CompAlloc}^{i}_{j}$, \textcolor{black}{\# of component $j$ for layer $i$} & No     & No    \\ 
          \hline
          \end{tabular}
          \end{table}
        
      \textcolor{black}{To comprehensively optimize PIM-based CNN accelerators}, 
        PIMSYN integrates an architectural exploration flow in the synthesis framework. As shown in Fig.~\ref{fig:overview}, \textcolor{black}{DSE is realized by iteratively exploring through the four synthesis stages}. In each iteration, each stage provides a set of design variables for exploration. \textcolor{black}{These design variables collectively determine the generated CNN accelerators.}
        Through continuous iterations and performance evaluations, PIMSYN finds accelerator implementations with optimal power efficiency.

        The design space is summarized in Table~\ref{tab:dse}. Compared with existing architecture exploration works~\cite{gibbon2,nacim}, we expand the design space in several ways. For example, ${RatioRram}$ and $\bm {CompAlloc}$\footnote{A bold variable denotes a vector containing the corresponding elements of all layers.} reflect the power distribution between different resources and different layers, which greatly affect the accelerator performance. In existing works, they are manually determined. The scale of our defined design space can reach up to {}{$10^{27}$} for VGG13~\cite{vgg}, making it impossible to traverse all cases. We adopt a simulated annealing (SA) process and an evolution algorithm (EA) to improve the exploration efficiency (see Section~\ref{pimsyn}). Algorithm~\ref{alg:dse} shows the DSE flow embedded in the synthesis framework. It is a multi-loop based  process to iteratively search for the optimal design variables listed in Table~\ref{tab:dse}, with embedded SA and EA for boosting the DSE flow.

       \begin{algorithm}[t]
\caption{Design space exploration flow.}
\scriptsize
\label{alg:dse}
\KwOut{Accelerator implementation} 
$Best\_arch \leftarrow$ NULL\;
$Best\_perf \leftarrow 0$\; 
\For{$RatioRram\in [0.1, 0.4]$}{
    \For{$ResRram \in \{1, 2, 4\}$}{
    \For{$XbSize \in \{128, 256, 512\}$}{
        $WtDupCandi \leftarrow$ top 30 solutions of SA-based filter\;
        \For{$\bm{WtDup} \in WtDupCandi$}{
        \For{$ResDAC \in \{1, 2, 4\}$}{
        Generate IR-based dataflow DAG\;
        $\bm{MacAlloc}, \bm{CompAlloc}$ $\leftarrow$ best solution of EA-based macro partitioning together with components allocation\;
        Evaluate performance of currently found best solution\;
        Update $Best\_perf$ and $Best\_arch$\;
        }
    }
    }
}
}

\end{algorithm}

\section{PIMSYN Synthesis Stages}\label{pimsyn}

\subsection{Weight Duplication}
 
\subsubsection{Problem Definition}
    Duplicating a layer's weights to more crossbar sets increases parallelism. Different layers are executed in a pipelined way on a PIM accelerator, so limited crossbars should be properly distributed to all layers. The distribution strategy decides weight duplication for each layer and  impacts the overall performance. We formulate the decision of weight duplication as a constrained optimization problem as
   
    \begin{equation}\label{eq:wtdup}
    \footnotesize
    \begin{aligned}
    & \underset{\bm{WtDup}}{\text{maximize}}& & Performance(\bm{WtDup}) \\
    & \text{subject to}& & \sum\nolimits_{i=1}^{L}\left(WtDup^{i}\times set^{i}\right) \leq \# crossbar.
    \end{aligned}
    \end{equation}
    We use $\bm {WtDup}$ to symbolize the weight duplication strategy. ${L}$ is the number of layers in the CNN model. $WtDup^{i}\times set^{i}$ is the number of crossbars used for layer $i$, where $set^{i}$ is calculated by Eq.~\eqref{eq:set}, which depends on two design variables $XbSize$ and $ResRram$.
    
    Solving Eq.~\eqref{eq:wtdup} also depends on $\#crossbar$ (total number of crossbars) 
    $\#crossbar$ can be determined by the user given total power constraint ($TotalPower$), $RatioRram$, and the power of a single crossbar that depends on $XbSize$ and $ResRram$. 
    As described in Fig.~\ref{fig:overview}, $\#crossbar$ is specified by three  design variables, ${RatioRram}$, ${XbSize}$ and  ${ResRram}$: 
    \begin{equation}
    \footnotesize
    \#crossbar = \frac{TotalPower\times RatioRram}{{CrossbarPower(XbSize, ResRram)}}.
    \end{equation}\label{eq:crossbar}
    
    The three design variables are explored through traversing them in the PIM-related design space, as shown in lines 3-5 in Alg.~\ref{alg:dse}. They provide  ${\#crossbar}$ for the weight duplication stage and affect the subsequent synthesis stages. The design space of ${RatioRram}$, ranging from 0.1 to 0.4, is derived from our prior knowledge. 

\subsubsection{SA-based Weight Duplication Filter}
    The size of $\bm{WtDup}$'s exploration space is the number of all possible positive integer solutions of the problem of Eq.~\eqref{eq:wtdup}. It is typically an astronomical search space. Since $\bm {WtDup}$ significantly influences the accelerator's performance, solutions that underperform in this stage are unlikely to become the final optimal solutions found by the DSE flow, which offers the opportunity for design space pruning. To shorten the DSE time, we integrate an SA-based filter in this stage to select 30 weight duplication candidates with the {lowest} energy-function values (line 6), which will be traversed in the future synthesis stages (line 7). Given a specific power constraint, the energy-function of SA is the accelerator's performance.  Nonetheless, precisely assessing the accelerator's performance needs the incorporation of subsequent synthesis stages, which complicates the synthesis process. Instead, we devise a straightforward energy-function to represent the performance, which is
    \begin{equation}\label{eq:sa}
    \footnotesize
        \begin{aligned}
           & EnergySA \! =\! \mathop{\rm stdev}\limits_{i=1,\!\cdots,\!L}\!\left(\frac{W_{O}^{i}H_{O}^i}{WtDup^{i}}\right) + \alpha\!\cdot\! \!\mathop{\rm stdev}\limits_{i=1,\!\cdots,\!L}\!(AccessVolume^{i}) \\
           & AccessVolume^{i} = WtDup^{i}\times (W_{K}^{i}\times W_{K}^{i}\times C_{I}^{i} + C_{O}^{i}),
        \end{aligned}
    \end{equation}
where $\alpha$ is an empirical parameter. $AccessVolume^{i}$ is the data access volume of layer $i$. Performance-optimal accelerators need to balance computation workload and data access latency for each layer. The SA-based filter selects weight duplication candidates with the smallest ${EnergySA}$ values that try to balance computation and data access for each layer.

\begin{table*}[t]
    \caption{List of intermediate representations.}\label{tab:ir}\scriptsize
    \centering
    \begin{threeparttable}
        \begin{tabular}{|c|c|p{4cm}|p{7.5cm}|}
        \hline
       Category      & IR       & \multicolumn{1}{c|}{Parameters}      & \multicolumn{1}{c|}{Parameter Explanation}    \\ 
        \hline
        \multirow{3}{*}{Computation}& MVM\tnote{a}& layer, cnt, bit, xb\_num & cnt: which computation block is currently being calculated                 
        \\ \cline{2-3}
        & ADC      & layer, cnt, bit, vec\_width  & layer: which layer the IR belongs to    \\ 
        \cline{2-3}
         &ALU &aluop, layer, cnt, bit, vec\_width   & bit: which bit is currently being calculated in computation block \\ 
        \cline{1-3} 
        \multirow{2}{*}{\begin{tabular}[c]{@{}c@{}}Intra-Macro\\ Communication\end{tabular}} & load     & layer, cnt, vec\_width             & xb\_num: number of crossbars allocated to the layer   \\ 
       \cline{2-3} 
       & store    & layer, cnt, vec\_width      & vec\_width: length of the operand     \\ 
       \cline{1-3} 
        \multirow{2}{*}{\begin{tabular}[c]{@{}c@{}}Inter-Macro\\ Communication\end{tabular}} & merge    & layer, macro\_num, vec\_width          & aluop: vector arithmetic/logical/non-linear operations  \\ 
       \cline{2-3} 
    & transfer & layer, src, dst, vec\_width  & macro\_num: number of macros partitioned to the layer        \\ 
    \hline
        \end{tabular}
       \begin{tablenotes}
           \item[a] MVM involves DAC and sample-hold. Due to the analog properties, the three operations cannot be divided into different control steps.
           \end{tablenotes}
           \end{threeparttable}
\end{table*}

\subsection{Dataflow Compilation}

This stage translates the CNN structural description into an IR-based dataflow DAG (line 9), which provides a unified representation of the CNN for the subsequent synthesis stages.  The compilation needs the weight duplication strategy from the previous stage and $ResDAC$ as inputs. 
In the IR-based DAG, nodes represent operations and edges depict the dependencies between operations. 
 IR acts as the interface between high-level algorithms and low-level implementations. Table~\ref{tab:ir} lists the defined IRs, including three categories: computation, intra-macro communication and inter-macro communication. The dataflow compilation has three steps. 


First, we translate each layer's computation into a set of IRs. {}{As depicted in Fig.~\ref{fig:cnn} and Section~\ref{sec:base}, in PIM-based accelerators, parallelism exists in two levels: computation block level and input bit level.} Any computation operation of a layer can be denoted by three indices:  \textit{layer}, \textit{cnt} and \textit{bit} (see Table~\ref{tab:ir} for the meanings of the notations). In PIMSYN, we complement the computation IRs with the three parameters as shown in Table~\ref{tab:ir}. We can specify the detailed operations of each layer, when $\bm {WtDup}$ and  ${ResDAC}$ are provided.

\begin{figure}[t]
    \centering
    \includegraphics[width=\columnwidth]{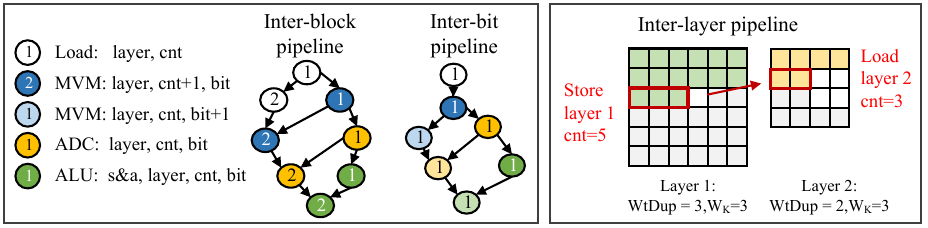}\\
    \caption{Dependency relationship between IRs.}\label{fig:compiler}
\end{figure}

Second, We establish the dependencies between IRs, including inter-layer, inter-block, inter-bit, and inter-operation dependencies, as shown in Fig.~\ref{fig:compiler}. Inter-layer dependency is decided by a fine-grained pipeline, which means that a layer can start computation as soon as the previous layer has produced sufficient outputs.
Computations of different \textit{computation blocks} and different bits are also performed in a pipelined manner. Inter-operation dependency indicates the order of operations executed within a \textit{computation block}.

Finally, we generate the IR-based DAG. Different from PUMA~\cite{puma}, our IRs can intuitively reflect the computing process of each layer as well as the status of the inter-layer pipeline, which specify the dataflow scheduling.  As each IR corresponds to a specific hardware intrinsic, hardware exploration is equivalent to find the optimal resource allocation for IRs. As a result, the performance of synthesized accelerators can be estimated by the depth of the IR-based DAG and the IRs' latencies, which will be discussed in Section~\ref{compalloc}.

\subsection{Macro Partitioning}

The high parallelism of PIM-based accelerators brings tremendous pressure to communication. If the resources of a layer are concentrated within a single macro, the communication overhead may be intolerable.  Taking ISAAC~\cite{ISAAC} accelerating conv3 of VGG16 as an example, the weight duplication of conv3 is 64, so the amount of inputs loaded once is about 64KB, taking 20 cycles. To alleviate the stress, we can partition a layer's resources into multiple macros, and they exchange data through the NoC. However, as the number of macros increases, communication between macros becomes complicated, introducing new challenges. The increase in macros also leads to a higher demand for storage resources and peripheral components, especially ADCs. In this stage, we implement an EA-based explorer to search for the macro partitioning solution with the optimal performance, denoted as $\bm{MacAlloc}$. This stage further supplements communication-related IRs to the dataflow DAG.
\subsubsection{Rules of Macro Partitioning}\label{sec:macro_1}

\begin{figure}[!t]
    \centering
    \includegraphics[width=.7\columnwidth]{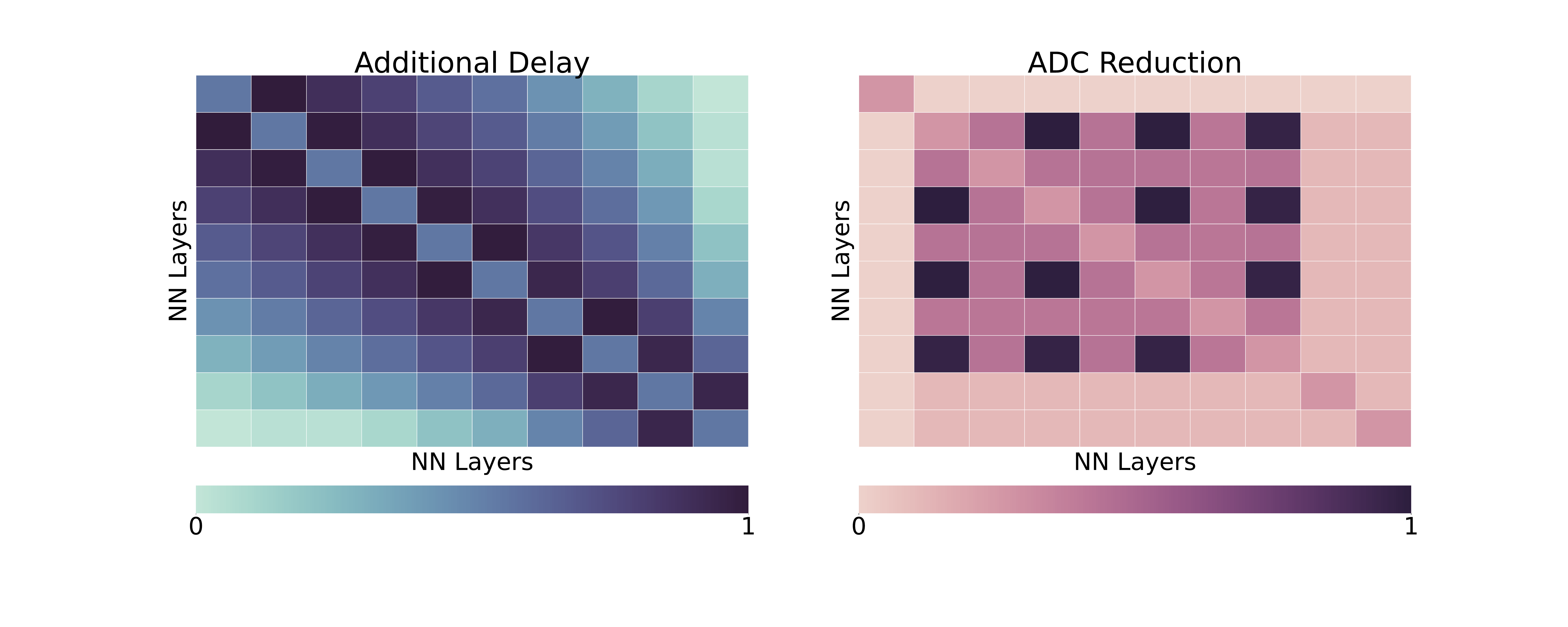}\\
    \caption{(a) Normalized delay caused by inter-layer ADC reuse. (b) Normalized number of reduced ADCs after reuse.}\label{fig:adc_reuse}
\end{figure}

PIMSYN supports two scenarios of macro implementations: identical macros across all layers and customized macros for different layers. We stipulate several rules to limit the feasible exploration space of macro partitioning.\\
    a) A layer can occupy one or more macros.\\
    b) Two layers can share the same set of macros.\\ 
    c) Layer $i$'s crossbars can be partitioned to at most $WtDup^{i}\times \frac{W_{K}^{i}W_{K}^{i}C_{I}^{i}}{XbSize}$ macros (a macro must have at least one crossbar).\\
\textcolor{black}{We introduce the opportunity of two layers sharing the same set of macros (rule b).}
Through macro sharing, two layers reuse the peripheral components, especially ADCs, at different times. We add the macro partitioning choice inspired by the following observation. In PIM-based CNN accelerators, ADCs consume over 60$\%$ of total power~\cite{ISAAC}. To reduce ADC power, most studies adopt intra-layer ADC reuse among columns of a crossbar~\cite{ISAAC}. However, we find that different layers can stagger their times for using ADCs, which offers the possibility of inter-layer ADC reuse. Fig.~\ref{fig:adc_reuse} shows ADC reuse between different layers in an example accelerator. When two layers are relatively far apart, ADC reuse hardly brings delay penalty, but it decreases the requirement of ADCs, \textcolor{black}{which can improve power efficiency potentially.} Therefore, through searching for the suitable pairs of macro-sharing layers, power efficiency of accelerators can be further optimized.

Although the aforementioned rules can prune part of the $\bm{MacAlloc}$ design space, it is still impossible to traverse all the design choices. To address that, we propose an EA-based explorer for improving the DSE efficiency.

\subsubsection{EA-based Macro Partitioning Explorer}

In EA, a gene represents a macro partitioning candidate. $\bm{MacAlloc}$ is encoded by an integer vector, where $MacAlloc^{i}$ is denoted by $i\times 1000+\#macro^i$. If layers $j$ and  $i$ share the same macros ($j<i$), $MacAlloc^{i}$ is changed to $j\times 1000+\#macro^i$. Alg.~\ref{ea} shows the EA-based explorer, executed on line 10 of Alg.~\ref{alg:dse}.

\begin{algorithm}[t]
\scriptsize
\caption{EA-based macro partitioning explorer.}
\label{ea}
\KwOut{Best macro partitioning solution}
Randomly initialize the $\bm{MacAlloc}$ population\;
Evaluate each gene\;
\While{$iter<MaxEAIterations$}{
  Select parents based on their fitness\;
  Apply mutation related to \#macros of each layer\;
  Apply mutation related to macro-sharing\;
  Evaluate new children\;
  Insert the children to population\;
}
Return the best solution found\;
\end{algorithm}

To efficiently explore macro partitioning, we present two mutation mechanisms in EA: \textit{mutate\_num} is to mutate the number of macros assigned to each layer, and  \textit{mutate\_share} is to change the status of macro-sharing between layers. During the mutation, the generated children always obey the defined rules (Section~\ref{sec:macro_1}). 

We use accelerator performance as the fitness function in the EA-based explorer. As shown in Fig.~\ref{fig:overview}, in each EA iteration, the generated children are passed to the components allocation stage, which will fulfill the architecture implementation and return the performance evaluation result to the EA-based explorer. After multiple iterations, we find a $\bm{MacAlloc}$ that optimizes the accelerator performance. Meanwhile, $\bm{MacAlloc}$ specifies the latencies of the communication-related IRs and further completes the IR-based DAG.

\subsection{Components Allocation}\label{compalloc}

This stage accomplishes mapping between IRs and hardware resources. To maximize power efficiency, we need to optimally distribute power among IRs. Different IRs can be assigned to the same physical resource if and only if each pair of these IRs does not have usage conflict. 
Resource allocation for the \textit{MVM} IR and communication-related IRs are determined before. We develop a heuristic to determine the peripheral components allocation for solving $\bm{CompAlloc}$.

Peripherals are components outside the PE array, such as the ADC bank and various ALU units (see Fig.~\ref{fig:arch}), which consume a large proportion of total power. The latency of each IR is the ratio of the IR's workload to the amount of its assigned resources. As the pipeline period depends on the most time-consuming step, 
 we model the resource allocation problem as
\begin{equation}\label{eq:1}
\footnotesize
   \begin{aligned}
        & \min  \max_{\substack{1\leq i \leq L,\\ c\in components}} \frac{Wl^{i}_{c}}{Freq_{c}\cdot CompAlloc^{i}_{c}},
        \\
        &\text{subject to} \  \sum_{i=1}^{L}\sum_{{c\in components}} P_{c}\cdot CompAlloc^{i}_{c}  \\
        &~~~~~~~~~~~~= (1-RatioRram)\times TotalPower,
    \end{aligned}
\end{equation}
where $CompAlloc_{c}^{i}$ is the number of functional unit $c$ allocated to layer $i$. $Wl^{i}_{c}$ is to the amount of workload that component $c$ needs to perform for layer $i$. $Freq_{c}$ and $P_{c}$ are the frequency and power of component $c$, respectively. Eq.~\eqref{eq:1} minimizes the maximum delay under the power limit of the functional units. It is obvious that the best components allocation should balance the delays of all steps. Thus, the solution is ($\forall l \in \{1,\cdots,L\}$ and $\forall \textcolor{black}{p} \in components$)
\begin{equation}\label{eq:2}
\footnotesize
\begin{aligned}
& (CompAlloc^{l}_{\textcolor{black}{p}})_{opt}  \times \sum_{i=1}^{L}\sum_{c\in {components}} \frac{P_{c}\times Wl^{i}_{c}}{Freq_{c}} \\
 & = (1-RatioRram)\times TotalPower\times\frac{Wl^{l}_{\textcolor{black}{p}}}{Freq_{\textcolor{black}{p}}}.
\end{aligned}
\end{equation}

\section{Experimental Results}

\begin{table}[t]
  \centering
  \caption{Part of evaluation and exploration parameters.}\label{tab:para}\scriptsize
  \begin{tabular}{|l|l|l|l|}
  \hline
  Component       & Parameters        & Values     & Power           \\ \hline\hline
  eDRAM         & size, bus\_width   & 64KB, 256b  & 20.7mW           \\ \hline
  NoC          & flit\_size, num\_port & 32, 8     & 42mW            \\ \hline
  \multirow{2}{*}{ReRAM} & crossbar size         & 128, 256, 512 & \multirow{2}{*}{0.3-4.8mW} \\ \cline{2-3}
             & precision       & 1, 2, 4    &              \\ \hline
  DAC          & resolution      & 1, 2, 4    & 4-30$\mu$W         \\ \hline
  ADC          & resolution      & 7, 8, $\cdots$, 14 & 2-54mW           \\ \hline
  \end{tabular}
  \end{table}

We compare the auto-synthsized accelerators with five manually-designed PIM accelerators~\cite{ISAAC,pipelayer,prime,puma,atomlayer}. 
We further compare PIMSYN with the recent model and architectural co-exploration work, Gibbon~\cite{gibbon2}. The benchmarks include AlexNet~\cite{alexnet}, VGG13~\cite{vgg}, VGG16~\cite{vgg}, MSRA~\cite{msra} and ResNet18~\cite{resnet} with 16-bit quantification. 
Some setup parameters in PIMSYN are listed in Table~\ref{tab:para}, together with some exploration parameters during DSE (defined in Table~\ref{tab:dse}).  Table~\ref{tab:para} only lists key parameters, and other parameters are provided by ISAAC~\cite{ISAAC} and MNSIM~\cite{MNSIM}. 
The synthesized accelerators are evaluated by a cycle-accurate  IR-based behavior-level simulator. 
PIMSYN is implemented in Python and it takes about 4 hours  to complete a synthesis process.

\subsection{Comparisons with Manually-Designed Architectures}
\textbf{Peak Power Efficiency.} As shown in Table~\ref{tab:peak}, PIMSYN achieves 3.65-21.45$\times$ peak power efficiency improvements, 8.11$\times$ on average, compared with the five state-of-the-art PIM-based CNN accelerators. 

\begin{table}[t]
    \centering
    \caption{Peak power efficiency comparison.}\label{tab:peak}\scriptsize
    \setlength{\tabcolsep}{2pt}
    \begin{threeparttable}
    \begin{tabular}{ccccccc}
      \toprule
      {} & {PIMSYN} & {PipeLayer} & {ISAAC} & {PRIME\tnote{a}} & {PUMA} & {Atomlayer}  \\
      {} & {} & {\cite{pipelayer}} & {\cite{ISAAC}} & {\cite{prime}} & {\cite{puma}} & {\cite{atomlayer}} \\
      \midrule
      {TOPS/W} & {3.07} & {0.14} & {0.63} & {0.5} & {0.84} & {0.68} \\
      {Improvement}   & {}      & {21.45$\times$}        & {4.83$\times$}  & {6.11$\times$} & {3.65$\times$} & {4.51$\times$}   \\
      \bottomrule
    \end{tabular}
    \begin{tablenotes}
    \item[a] Projected to 16-bit quantification as PRIME uses 8-bit quantification.
    \end{tablenotes}
    \end{threeparttable}
\end{table}

\begin{figure}[!t]
    \centering
    {
      \includegraphics[width=.4\columnwidth]{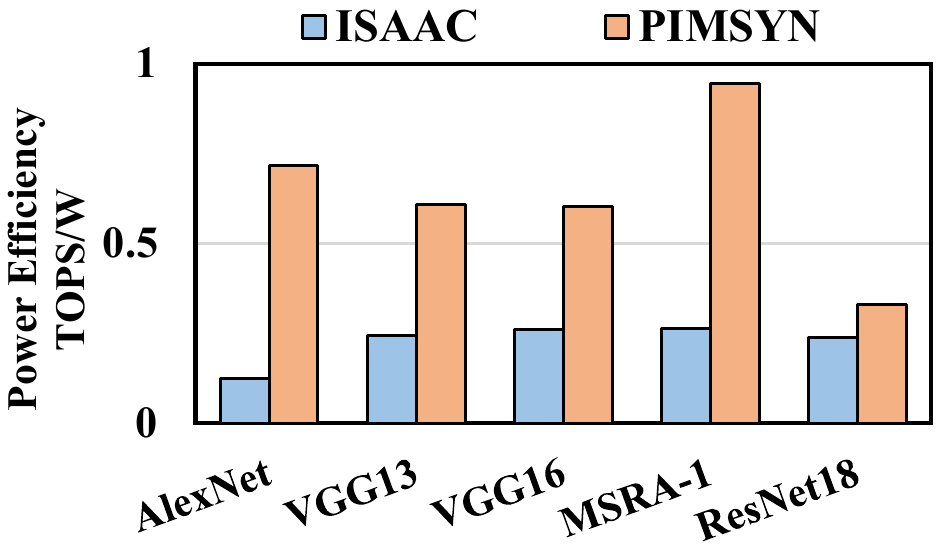}\label{fig:isaac}
    }
    {\label{fig:matr}
      \includegraphics[width=.4\columnwidth]{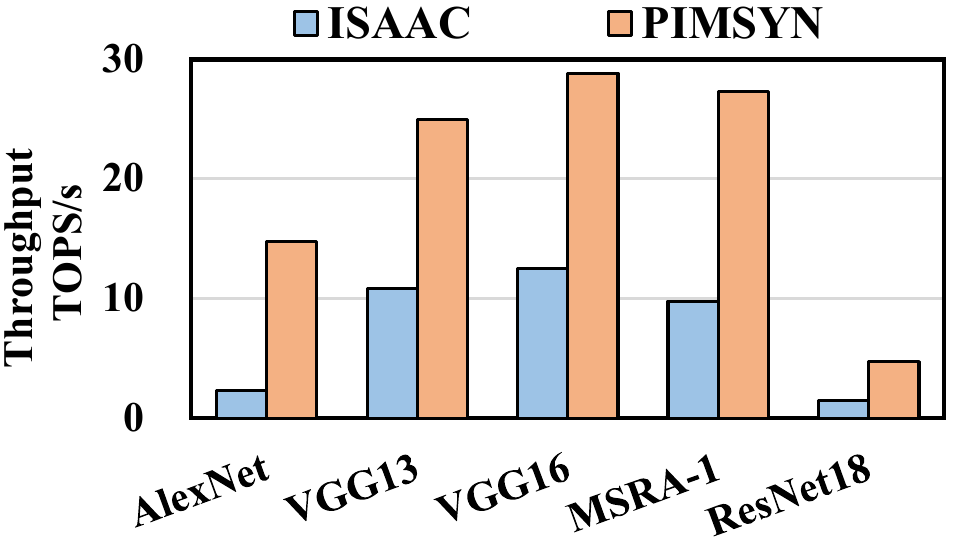}\label{fig:throughput}
    }
    \caption{Effective power efficiency and throughput comparisons with ISAAC.}\label{fig:6}
  \end{figure}

\textbf{Effective Power Efficiency.} 
It means the achievable power efficiency when running a specific CNN model. Here only ISAAC~\cite{ISAAC} is compared, because only ISAAC offers detailed parameters to assess the effective power efficiency. 
Fig.~\ref{fig:6} shows that PIMSYN outperforms ISAAC for all five models, with 1.4-5.8$\times$ improvements in power efficiency (3.9$\times$ on average). The improvement comes from the better power distribution among hardware components. ISAAC has a large portion of power  ($>$80\%) consumed by peripheral components, decreasing crossbars' power and computation parallelism.

\textbf{Throughput.} As shown in Fig.~\ref{fig:6}, PIMSYN achieves 2.30-6.45$\times$ (3.4$\times$ on average) higher throughput than ISAAC. The reason of the higher throughput is the same as that of the power efficiency improvement.

\subsection{Comparisons with Architectural Exploration Work}

 Here we compare PIMSYN with Gibbon~\cite{gibbon2}, which fulfills a co-exploration flow of CNN models and architectures.  As PIMSYN does not involve model exploration, we take CNN models which are trained by Gibbon as inputs. 
    Table~\ref{tab:cifar10} shows that PIMSYN finds better architectures with an average of 56\% decrease on energy-delay product (EDP), compared with Gibbon, for CIFAR-10/CIFAR-100 datesets.

\begin{table}[!t]
\centering
\caption{Comparisons with Gibbon for CIFAR-10/CIFAR-100 datasets.}\label{tab:cifar10}\scriptsize
\begin{tabular}{|cc|c|c|c|}
\hline
\multicolumn{2}{|c|}{}        & EDP (ms$\times$mJ) & Energy (mJ) & Latency (ms) \\ \hline
\multicolumn{1}{|c|}{\multirow{2}{*}{AlexNet}}  & Gibbon & 0.38/0.38   & 0.38/0.38  &  0.99/0.99                                                                           \\ \cline{2-5} 
\multicolumn{1}{|c|}{}                          & PIMSYN    &    0.024/0.024                       & 0.119/0.119 &  0.197/0.197                                                                           \\ \hline
\multicolumn{1}{|c|}{\multirow{2}{*}{VGG16}}    & Gibbon & 17.22/17.25  &      2.68/2.68      &  6.43/6.44                                                                           \\ \cline{2-5} 
\multicolumn{1}{|c|}{}  & PIMSYN      &  7.94/7.95 & 2.98/2.99 & 2.66/2.66                                                                            \\ \hline
\multicolumn{1}{|c|}{\multirow{2}{*}{ResNet18}} & Gibbon&  4.75/4.76 & 1.33/1.33  &  3.58/3.58                                                                           \\ \cline{2-5} 
\multicolumn{1}{|c|}{}  & PIMSYN    &   3.76/3.78    &  2.34/2.35   &   1.61/1.61                                  \                                     \\ \hline
\end{tabular}
\end{table}

\begin{figure}[!t]
    \centering
    {
      \includegraphics[width=.4\columnwidth]{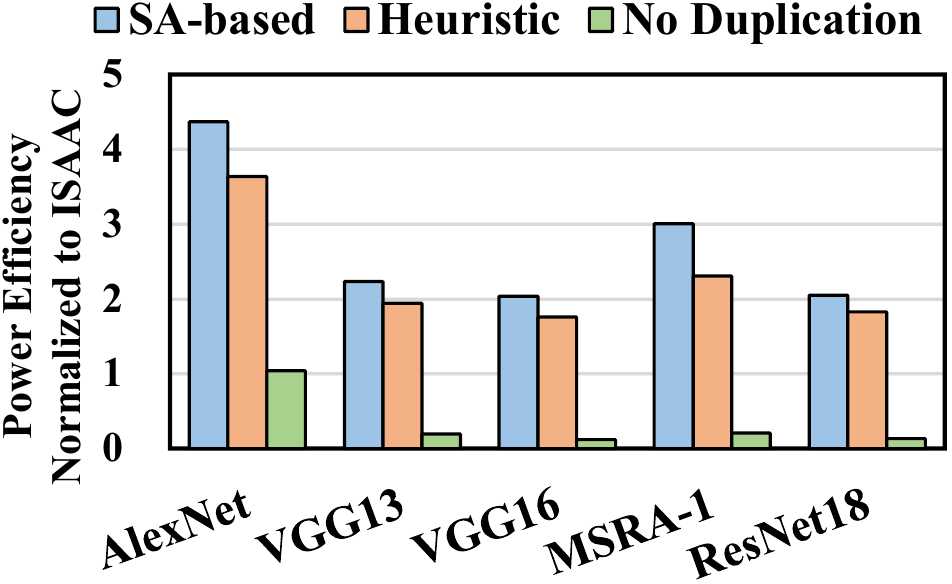}
    }
    {\label{fig:matr}
      \includegraphics[width=.4\columnwidth]{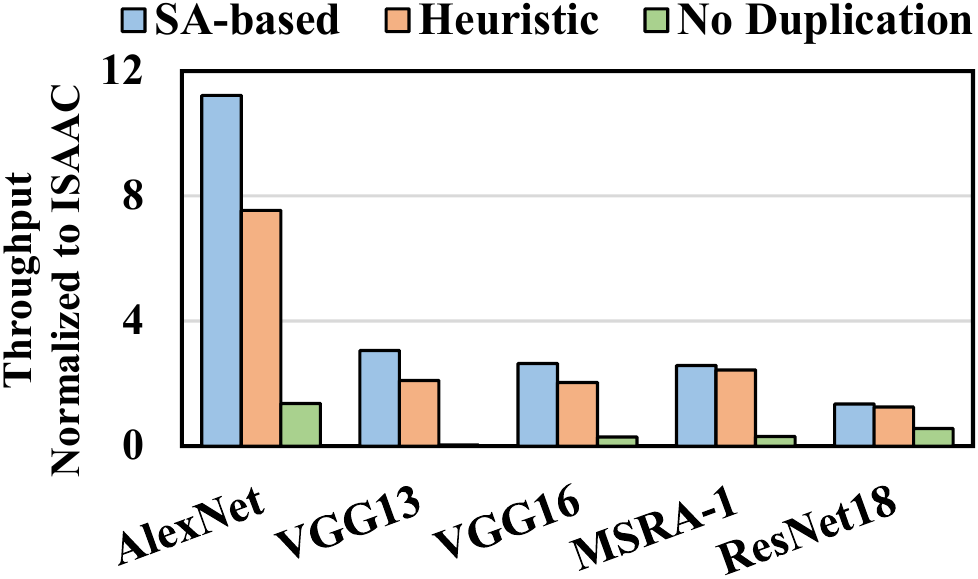}
    }
    \caption{Power efficiency and throughput improvements brought by different weight duplication methods.}\label{fig:sa}
  \end{figure}

\subsection{Effectiveness of Enlarged Design Space}

Here  we demonstrate the effectiveness of the enlarged design space in PIMSYN, as listed in Table~\ref{tab:dse}.

\subsubsection{SA Selected Weight Duplication}
In PIMSYN, $\bm{WtDup}$ is selected by the SA-based filter. Previous works~\cite{ISAAC, pipelayer} use a heuristic \bm{$W_{O}H_{O}$}-proportional  method (layers' weight duplication factors  are proportional to layers' $W_O H_O$).  Fig.~\ref{fig:sa} compares the two methods. 
 PIMSYN achieves 19\% power efficiency improvement and 27\% throughput improvement. The heuristic  decides $\bm{WtDup}$ based on layers' workloads,
 which intuitively tries to balance layers' computation latencies. However,  it usually requires a large number of crossbars and sufficient bandwidth to support high computation parallelism. In contrast, PIMSYN can be applied to various scenarios, even for power-constrained and bandwidth-limited conditions. 

Existing architectural exploration works~\cite{gibbon2,nacim} do not involve weight duplication. In such a case, the power efficiency and throughput are tens of times lower (Fig.~\ref{fig:sa}), indicating the necessity of weight duplication in the designs of PIM-based CNN accelerators and the corresponding workload mapping.

\subsubsection{Specialized Macro Design}
\begin{figure}[!t]
  \centering
  {
    \includegraphics[width=.4\columnwidth]{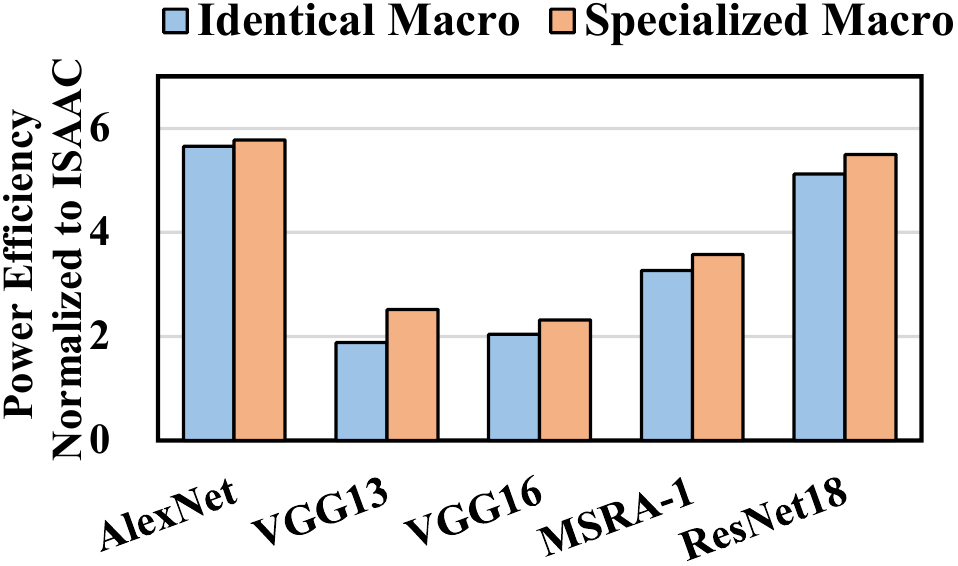}
  }
  {\label{fig:matr}
    \includegraphics[width=.4\columnwidth]{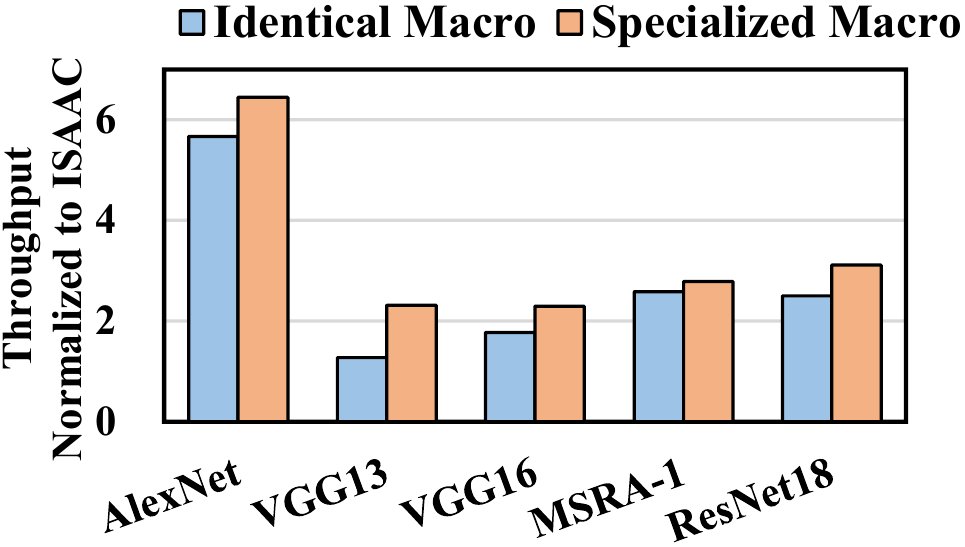}
  }
  \caption{Power efficiency and throughput improvements brought by specialized macro design.}\label{fig:macro}
\end{figure}

The above results are all for specialized macros. Fig.~\ref{fig:macro} compares between synthesized accelerators with identical and specialized macros. 
It show that the specialized macro design realizes 13\% power efficiency improvement and  31\% throughput improvement. The advantages come from the decrease of the macros' number, which  reduces the memory power and inter-macro communication. 

\begin{figure}[!t]
  \centering
  {
    \includegraphics[width=.4\columnwidth]{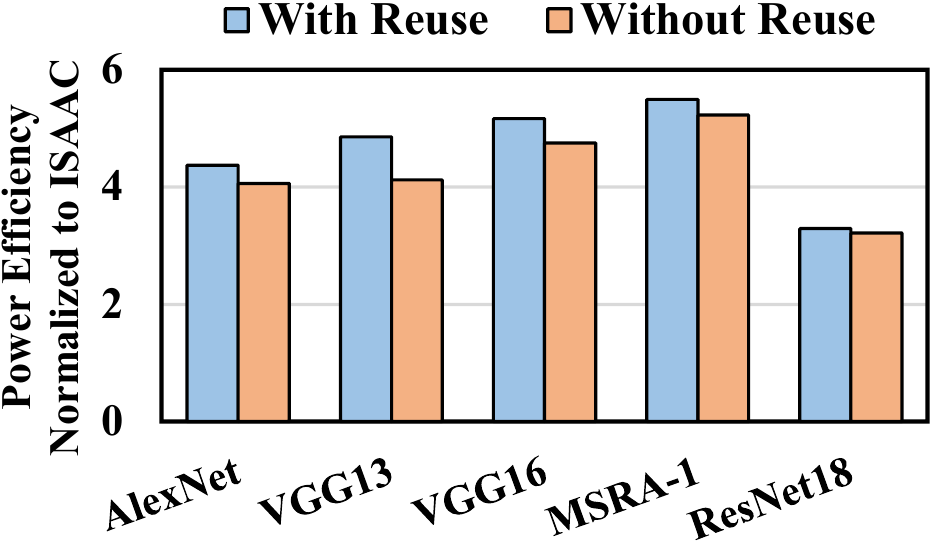}
  }
  {
    \includegraphics[width=.4\columnwidth]{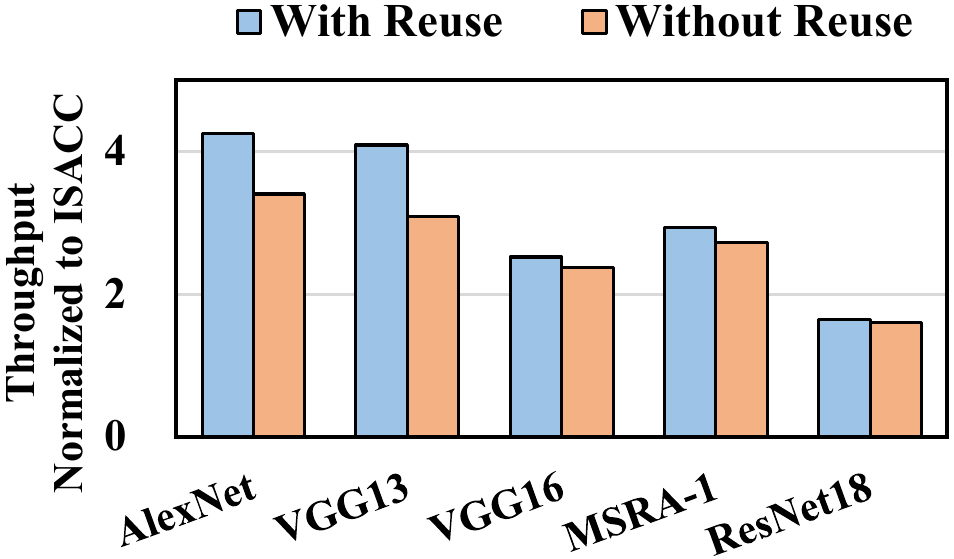}
  }
  \caption{Power efficiency and throughput improvements brought by inter-layer macro sharing.}\label{fig:reuse}
\end{figure}

\subsubsection{Inter-layer Macro Sharing}

Fig.~\ref{fig:reuse} compares between synthesized accelerators with and without inter-layer macro reuse. Macros in both cases are specialized.  Inter-layer ADC reuse makes the effective number of ADCs allocated to the shared layers more than before. It  shortens the inference time if the pipeline period is dominated by  ADCs. 
Inter-layer macro sharing increases 8\% and 15\% in the overall power efficiency and throughput, respectively.

\section{conclusion}

This work develops an automatic synthesis framework for generating PIM-based CNN accelerators, which frees architecture designers from the complexity of manual design. Given CNN models and power constraints, PIMSYN synthesizes the dataflows and architectures of  power-efficient accelerators. 
Compared with previous works, PIMSYN greatly enhances the potential of CNN acceleration by PIM.
\textcolor{black}{PIMSYN actually does not rely on the specific device, like ReRAMs. It uses the abstract architecture template that needs some device parameters (e.g., read power and latency). PIMSYN can be used to synthesize any crossbar-based PIM CNN accelerators.}

\bibliographystyle{IEEEtran}
\bibliography{my}

\end{document}